\documentclass[manuscript]{acmart}

\AtBeginDocument{%
  \providecommand\BibTeX{{%
    \normalfont B\kern-0.5em{\scshape i\kern-0.25em b}\kern-0.8em\TeX}}}

\setcopyright{rightsretained} 
\copyrightyear{2021}
\acmYear{2021}

\acmBooktitle{CHI Conference on Human Factors in Computing Systems Extended Abstracts (CHI '21 Extended Abstracts), May 8--13, 2021, Yokohama, Japan}

\usepackage{array}
\usepackage{arydshln}
\begin{document}

\title{How Social Are Social Media: The Dark Patterns In Facebook's Interface}


\author{Thomas Mildner}
\affiliation{%
  \institution{University of Bremen}
  \city{Bremen}
  \country{Germany}
}

\author{Gian-Luca Savino}
\affiliation{%
  \institution{University of Bremen}
  \city{Bremen}
  \country{Germany}
}

\renewcommand{\shortauthors}{Mildner, Savino}

\begin{abstract}
Many researchers have been concerned with social media and possible negative impacts on the well-being of their audience. With the popularity of social networking sites (SNS) steadily increasing, psychological and social sciences have shown great interest in their effects and consequences on humans. Unfortunately, it appears to be difficult to find correlations between SNS and the results of their works. We, therefore, investigate Facebook using the tools of HCI to find connections between interface features and the concerns raised by these domains. With a nod towards Dark Patterns, we use an empirical design analysis to identify interface interferences that impact users’ online privacy. We further discuss how HCI can help to work towards more ethical user interfaces in the future.
\end{abstract}

\ccsdesc[500]{Human-centered computing~Empirical studies in HCI}
\ccsdesc[100]{Human-centered computing~HCI theory, concepts and models}
\ccsdesc[300]{Human-centered computing~Empirical studies in interaction design}
\ccsdesc[100]{Human-centered computing~Interaction design theory, concepts and paradigms}
\ccsdesc[300]{Security and privacy~Usability in security and privacy}

\keywords{SNS, social media, Facebook, interface design, Dark Patterns, well-being, ethical interfaces}

\maketitle

\section{Introduction \& Motivation}
The growth of social networking sites (SNS) has evoked an interest in their effects on people among different research disciplines. For some time now, studies from psychological and social sciences have discussed how SNS may affect peoples' well-being with a particular focus on mental health. While some results show small positive effects when spending time on social media, others yield small negative effects, or did not find any effect at all. On the one hand, various studies have noted connections between increased depressive symptoms, that in occasions tragically resulted in suicides, and increased screen-time on SNS \cite{twenge2018}. On the other hand, researchers did not identify any such links \cite{coyne2020} or observed overall positive effects on people's well-being instead \cite{ahn2013social, grieve2013face, sinclair2017facebook}. 

These very different results urge for further research. A community with the tools to analyse and evaluate systems based on of their interface, HCI can add valuable insights to describe the relationship people have with social media. Understanding this relationship may reveal harmful design patterns that could affect well-being. Although participants of longitude studies self-reported a negative impact on physical and mental health when using Facebook  \cite{shakya2016intimate}, they still kept using the application regularly. Offering an explanation, Christakis highlights a link to possible internet addiction that causes users of Facebook to experience decreased future well-being \cite{christakis2009trapped, christakis2010internet}. 

Recent works in the area of HCI have already begun to research on SNS. Looking at Facebook, Wang et al. found that people often regret sharing certain content, which ultimately caused them disadvantages in their everyday lives \cite{wang2011}. In a follow-up study, they implemented additional features that give people the chance to review the content of their posts while displaying recipients before publishing \cite{wang2013}. Since most participants approved these features, they successfully demonstrated possibilities in which interface design can respond to people's concerns. Interface design often utilises knowledge about human psychology to create easy to use interactions. Although well-established usability is an achievement for designers, instances show that the same techniques could be used to misguide people into doing something that they either did not expect or even trick them into actions with harmful results. With regards to Brignull's work coining the term Dark Patterns \cite{brignull2015dark}, Grey et al. have shown such occurrences in various web-interfaces. 

At first sight, the implementation of Dark Patterns could be considered contradictions to core incentives of SNS to keep their audiences entertained while remaining credible themselves. As mostly free to use services, social media often rely on advertisement for revenue \cite{devries2012, kim2016}, which could create an interplay of the time people spend on a social medium and its generated income. Our research, therefore, investigates interface strategies of SNS that keep users active by implementing Dark Patterns or features similar to them. Based on empirical design analysis, we highlight scopes in which Facebook implements Dark Patterns, specifically interface interferences.


\section{Empirical Design Analysis}

\begin{figure}[t!]
    \centering
    \includegraphics[width=1\textwidth]{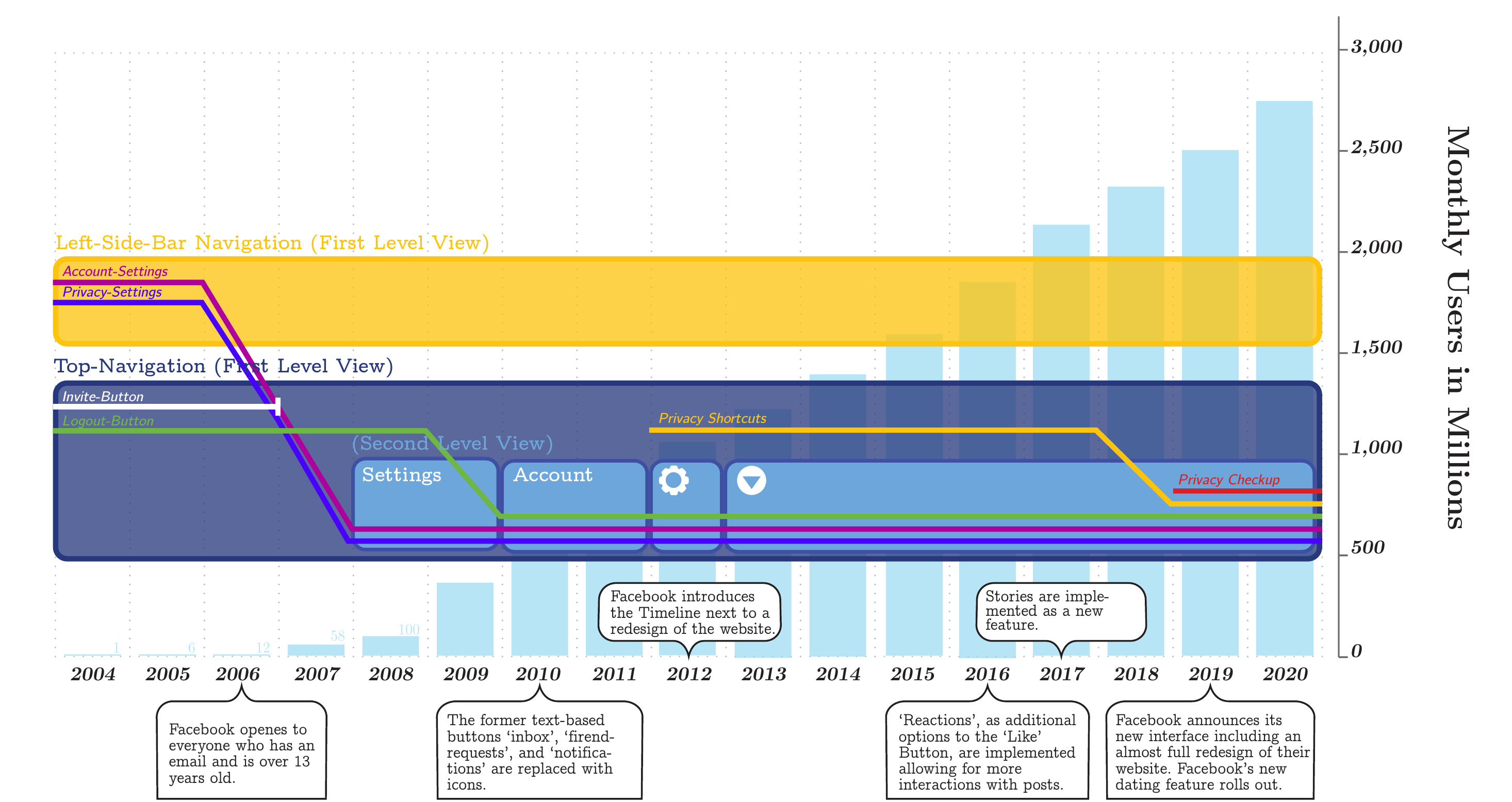}
    \Description[Figure 1 shows a diagram of Facebook’s interface changes between 2004 and 2020 with supplement data to highlight the increase of Facebook’s user count per year and descriptive texts for certain occasions.]
    {This diagram follows a time-axis from 2004 to 2020. Throughout the years, Facebook updated their user interface various times. The diagram describes interface elements as they were added to the application while following their position throughout multiple updates. Facebook’s first level views are mainly split into a side-bar and a top-navigation.
    In 2004, the side-bar included Account-Settings and Privacy-Settings. In 2008, these elements moved to the Top-Navigation into the second level-view of the settings menu. A descriptive text for 2006 reads: “Facebook opened to everyone who has email and is over 13 years old.” As a result of this change, an invite-button, located in the top-navigation since 2004, was soon removed in 2007. Also in the top-navigation, the logout button, which was also present since 2004, was moved into the second level view of the settings menu, now called ‘account’ menu.
    Another descriptive text for 2010 reads: “The former text-based buttons [for] ‘inbox’, ‘friend-requests’, and ‘notifications’ are replaced with icons”. 
    A descriptive text for 2012 states: “Facebook introduces the Timeline next to a redesign of the website” In this year, Facebook introduced another element – the Privacy Shortcuts. This feature was on the first-level view of the top-navigation until Facebook’s latest UI update in 2019. It was then moved into the ‘Account’ menu, now visualised by an icon of a downwards pointing arrow. This icon replaced a gear icon which was present in 2012. Also in 2019, Facebook introduced their Privacy Checkup next to the Privacy Shortcuts, a feature that allows users to make changes to pre-selected privacy settings.
    The diagram shows three  further texts: The first, a descriptive text for 2016 reads: “‘Reactions, as additional options to the ‘Like’ button, are implemented allowing for more interactions with posts.” The second text for 2017 reads: “Stories are implemented as a new Feature.” A last descriptive text for 2019 states: “Facebook announces its new interface including an almost full redesign of their website. Facebook’s new dating feature roll’s out.”

    For each year the respective user count is presented in a bar plot as supplement data. Facebook counted 1 Million users in their last quarter of 2004. In what looks like linear growth, the count increased to 2,740 Million users in the 3rd quarter of 2020. The user count for each year are as follows: 2004: 1; 2005: 6; 2006: 12; 2007: 58; 2008: 100; 2009: 360; 2010: 608; 2011: 845; 2012: 1056; 2013: 1228; 2014: 1393; 2015: 1591; 2016: 1860; 2017: 2129; 2018: 2320; 2019: 2498; 2020: 2740.
}
    \caption{This diagram visualises Facebook's interface changes between 2004-2020. With increasing popularity the interface was extended in multiple iterations. Because no official and complete data-set was present at the time of writing this work, the data of monthly Facebook users was collected from two sources. Firstly, The data for the years between 2004-2008 were retrieved from Facebook Newsroom \cite{facebook2021, sedghi2014}. As Facebook Newsroom no longer features their original data, an article from The Guardian that relies on the same source is cross-referenced for support. However, comparisons need to be tentative and taken with caution. Secondly, the data representing the years from 2008-2020 was gathered from the online service Statista \cite{clement2020}.}
    \label{fig:interfaceRecords}
\end{figure}

The goal of the empirical design study was to find indications for Dark Patterns within Facebook's interface. On his website, Brignull put out a list of various malicious interface patterns to create awareness among people. One of them is the term 'Privacy Zuckering', which he used to describe the phenomenon of people publicly sharing more about themselves than they wish to~\cite{brignull2015dark}. Motivated by this type of Dark Pattern, two HCI researchers analysed Facebook's desktop user-interface on an empirical level. To collect necessary screenshots of past UI iterations, we manually searched the web while comparing sources to ensure accuracy. Further analysis focused mainly on Dark Patterns and privacy concerns. It included tracking certain UI-elements like the logout button and its positioning in the navigation bar as well as several iterations of the 'account' menu. For readability, the most prominent findings are visualised in Figure \ref{fig:interfaceRecords}.

While most features remained constant elements, Facebook have continuously rearranged their interface. Especially the logout button and the privacy settings have received multiple changes. Both were moved from a first-level view into menus hindering their discoverability, which can be classified as interface interference~\cite{gray2018}. Even though these changes could be natural consequences of responsive interface design, it is important to note that Facebook benefits from users not logging out or sharing more information due to lighter privacy settings. This way, Facebook is able to track its users across the web and use the gathered data to create content and targeted advertisement. 

\begin{figure}[t!]
    \centering
    \includegraphics[width=1\textwidth]{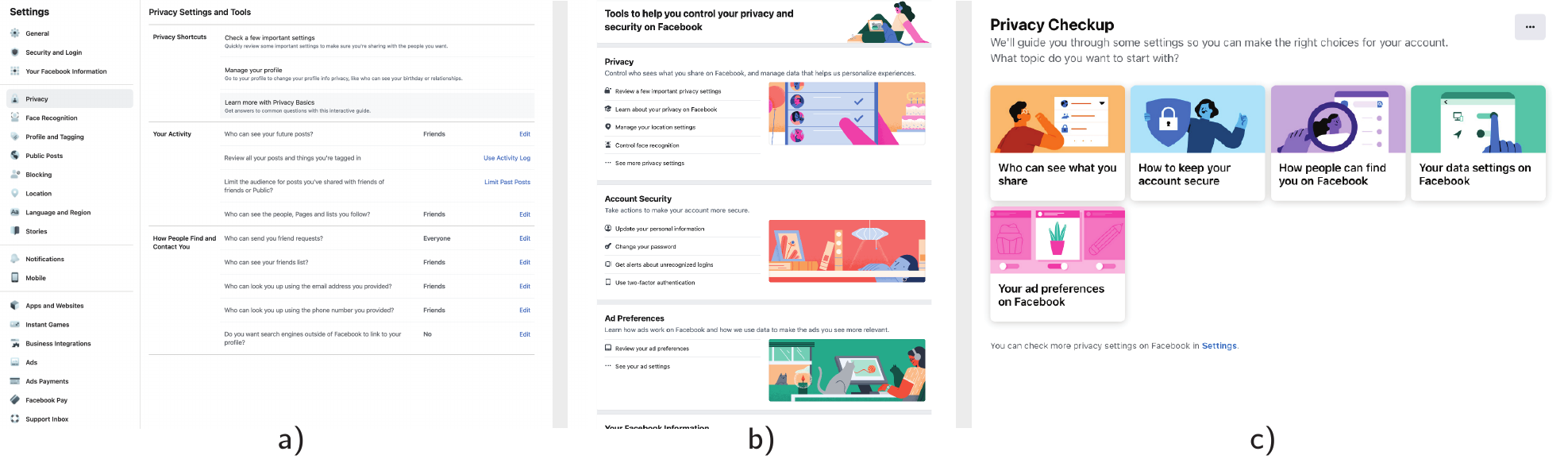}
    \Description[Figure 2 shows a comparison between Facebook’s general Privacy Settings, their Privacy Shortcuts, and their Privacy Checkup of Facebook's latest update.]
    {}
    \caption{Figure 2 shows a comparison between Facebook’s a) general Privacy Settings, their b) Privacy Shortcuts, and their c) Privacy Checkup of Facebook's latest update. Noticeably, the three interfaces yield different operability.}
    \label{fig:privacy-settings}
\end{figure}

The way, in which Facebook handles control over privacy settings sets an example for a novel Dark Pattern. As visualised in Figure \ref{fig:privacy-settings}, Facebook offers three different sections for users to engage with privacy options: (a) a general settings menu that includes all possible controls; (b) the Privacy Shortcuts, which yield links to the most used privacy settings; and (c) the Privacy Checkup, a fully interactive section that offers a guided interface to change selected privacy settings, albeit with incomplete coverage. Previous research has already shown that most people find it difficult to manage their privacy settings on Facebook, which often do not result in the expected level of security ~\cite{Liu2011}. Especially ad-related settings can interfere with Facebook's advertisement strategies. By offering a guided settings feature, Facebook is able to curate which settings users will manage. If this intentionally implemented to keep users from changing specific settings, it could be viewed as a novel form of interface interference. While placing all privacy settings behind several interface layers, Facebook actively offers a well designed but incomplete alternative to handle them.

\section{Discussion}
As so often in technology, SNS change quickly and thus require research to regularly re-evaluate their impact. HCI has the utilities to continuously analyse SNS and their effects on users with a particular focus on human-centred design. It can describe the relationship between a person and an interface and actively propose design guidelines to prevent possibly harmful design choices that affect people's well being. In 16 years of their existence, Facebook has continuously updated its interface while design choices have lead to potential Dark Patterns. The presented empirical design study identifies two interface interferences. As can be seen in Figure \ref{fig:interfaceRecords}, Facebook has changed the location of the logout button as well as the link to the privacy settings multiple times, while offering alternative but incomplete features. With both their Privacy Shortcuts and Privacy Checkups, Facebook provide their audience with alternative controls to the general privacy settings. By limiting the number of options, however, Facebook governs over the settings that their audience will adjust. Although such alternative interfaces are not covered under the existing Dark Patterns terminology, they create interferences that allow the navigation of people's decision-making without causing immediate harm. Governing but never harming users makes it possible to keep the general satisfaction of people using SNS high. 

Even though these interface changes may be reasoned through aesthetics, they result in unnecessary obstacles that people have to overcome and demonstrate possible areas for novel Dark Patterns in the future.  Observations of the presented interface analysis illustrate how the principles of Dark Patterns could be used to evaluate ethics in interface design by assessing the difficulty of, for example, basic user actions and evaluating peoples' expectations and motivations. The research on the protection of human's well-being on SNS is an interdisciplinary effort. HCI adds a range of utilities that help to understand the interaction between people and SNS on the level of interface and technology. This community can thus support psychological and social sciences by actively impacting future design development and providing ethical guidelines while advancing its interdisciplinary efforts. Future research could provide design practitioners with novel tools to avoid Dark Patterns and other unethical design strategies while empowering people by helping to understand and recognise their occurrences.

\section{Conclusion}
This work explores how analytical HCI methodologies can be used to find harmful design implementations in social networking sites such as Facebook. Through our interface analysis, we discovered Dark Patterns in the form of two interface interferences with regards to the logout button and the privacy settings. Furthermore, we discuss a possible novel Dark Pattern by highlighting Facebook's use of alternative controls to set selected privacy settings. By offering their audience a guided interface, Facebook governs over the options that their users will change while leaving them with a feeling of being in control. The worrying research of psychological and social scientists have motivated us to look at possible effects of SNS on well-being from the perspective of HCI and interface design. We suspect that future research should further observe individual people and their usage behaviour on social media more closely. As Beyens et al.~\cite{beyens2020effect} show, there can be vast individual differences, especially between adolescent users, in the way SNS affect their well-being. HCI methodologies, such as usability studies and interviews, could help to understand design mechanisms that cause negative effects on well-being. In this work, we came up with first guidelines for ethical interface design (e.g. testing for Dark Patterns) and we hope to spark the interest of the scientific community to engage in this research.

\bibliographystyle{ACM-Reference-Format}
\bibliography{references.bib}


\begin{thebibliography}{19}


\ifx \showCODEN    \undefined \def \showCODEN     #1{\unskip}     \fi
\ifx \showDOI      \undefined \def \showDOI       #1{#1}\fi
\ifx \showISBNx    \undefined \def \showISBNx     #1{\unskip}     \fi
\ifx \showISBNxiii \undefined \def \showISBNxiii  #1{\unskip}     \fi
\ifx \showISSN     \undefined \def \showISSN      #1{\unskip}     \fi
\ifx \showLCCN     \undefined \def \showLCCN      #1{\unskip}     \fi
\ifx \shownote     \undefined \def \shownote      #1{#1}          \fi
\ifx \showarticletitle \undefined \def \showarticletitle #1{#1}   \fi
\ifx \showURL      \undefined \def \showURL       {\relax}        \fi
\providecommand\bibfield[2]{#2}
\providecommand\bibinfo[2]{#2}
\providecommand\natexlab[1]{#1}
\providecommand\showeprint[2][]{arXiv:#2}

\bibitem[\protect\citeauthoryear{??}{fac}{2021}]%
        {facebook2021}
 \bibinfo{year}{2021}\natexlab{}.
\newblock \bibinfo{title}{Facebook Newsroom}.
\newblock
\newblock
\urldef\tempurl%
\url{https://about.fb.com/news/}
\showURL{%
\tempurl}
\newblock
\shownote{(visited on 01/05/2021).}


\bibitem[\protect\citeauthoryear{Ahn and Shin}{Ahn and Shin}{2013}]%
        {ahn2013social}
\bibfield{author}{\bibinfo{person}{Dohyun Ahn} {and} \bibinfo{person}{Dong-Hee
  Shin}.} \bibinfo{year}{2013}\natexlab{}.
\newblock \showarticletitle{Is the social use of media for seeking
  connectedness or for avoiding social isolation? Mechanisms underlying media
  use and subjective well-being}.
\newblock \bibinfo{journal}{\emph{Computers in Human Behavior}}
  \bibinfo{volume}{29}, \bibinfo{number}{6} (\bibinfo{year}{2013}),
  \bibinfo{pages}{2453--2462}.
\newblock


\bibitem[\protect\citeauthoryear{Beyens, Pouwels, van Driel, Keijsers, and
  Valkenburg}{Beyens et~al\mbox{.}}{2020}]%
        {beyens2020effect}
\bibfield{author}{\bibinfo{person}{Ine Beyens}, \bibinfo{person}{J~Loes
  Pouwels}, \bibinfo{person}{Irene~I van Driel}, \bibinfo{person}{Loes
  Keijsers}, {and} \bibinfo{person}{Patti~M Valkenburg}.}
  \bibinfo{year}{2020}\natexlab{}.
\newblock \showarticletitle{The effect of social media on well-being differs
  from adolescent to adolescent}.
\newblock \bibinfo{journal}{\emph{Scientific Reports}} \bibinfo{volume}{10},
  \bibinfo{number}{1} (\bibinfo{year}{2020}), \bibinfo{pages}{1--11}.
\newblock


\bibitem[\protect\citeauthoryear{Brignull, Miquel, Rosenberg, and
  Offer}{Brignull et~al\mbox{.}}{2015}]%
        {brignull2015dark}
\bibfield{author}{\bibinfo{person}{Harry Brignull}, \bibinfo{person}{Marc
  Miquel}, \bibinfo{person}{Jeremy Rosenberg}, {and} \bibinfo{person}{James
  Offer}.} \bibinfo{year}{2015}\natexlab{}.
\newblock \bibinfo{title}{Dark Patterns-User Interfaces Designed to Trick
  People}.
\newblock
\newblock
\urldef\tempurl%
\url{http://darkpatterns.org/}
\showURL{%
\tempurl}
\newblock
\shownote{(visited on 01/05/2021).}


\bibitem[\protect\citeauthoryear{Christakis}{Christakis}{2010}]%
        {christakis2010internet}
\bibfield{author}{\bibinfo{person}{Dimitri~A Christakis}.}
  \bibinfo{year}{2010}\natexlab{}.
\newblock \showarticletitle{Internet addiction: a 21 st century epidemic?}
\newblock \bibinfo{journal}{\emph{BMC medicine}} \bibinfo{volume}{8},
  \bibinfo{number}{1} (\bibinfo{year}{2010}), \bibinfo{pages}{1--3}.
\newblock


\bibitem[\protect\citeauthoryear{Christakis and Moreno}{Christakis and
  Moreno}{2009}]%
        {christakis2009trapped}
\bibfield{author}{\bibinfo{person}{Dimitri~A Christakis} {and}
  \bibinfo{person}{Megan~A Moreno}.} \bibinfo{year}{2009}\natexlab{}.
\newblock \showarticletitle{Trapped in the net: will internet addiction become
  a 21st-century epidemic?}
\newblock \bibinfo{journal}{\emph{Archives of pediatrics \& adolescent
  medicine}} \bibinfo{volume}{163}, \bibinfo{number}{10}
  (\bibinfo{year}{2009}), \bibinfo{pages}{959--960}.
\newblock


\bibitem[\protect\citeauthoryear{Clement}{Clement}{2020}]%
        {clement2020}
\bibfield{author}{\bibinfo{person}{J. Clement}.}
  \bibinfo{year}{2020}\natexlab{}.
\newblock \bibinfo{title}{Facebook MAU worldwide 2020}.
\newblock
\newblock
\urldef\tempurl%
\url{https://www.statista.com/statistics/264810/number-of-monthly-active-facebook-users-worldwide/}
\showURL{%
\tempurl}
\newblock
\shownote{(visited on 01/05/2021).}


\bibitem[\protect\citeauthoryear{Coyne, Rogers, Zurcher, Stockdale, and
  Booth}{Coyne et~al\mbox{.}}{2020}]%
        {coyne2020}
\bibfield{author}{\bibinfo{person}{Sarah~M. Coyne}, \bibinfo{person}{Adam~A.
  Rogers}, \bibinfo{person}{Jessica~D. Zurcher}, \bibinfo{person}{Laura
  Stockdale}, {and} \bibinfo{person}{McCall Booth}.}
  \bibinfo{year}{2020}\natexlab{}.
\newblock \showarticletitle{Does time spent using social media impact mental
  health?: An eight year longitudinal study}.
\newblock \bibinfo{journal}{\emph{Computers in Human Behavior}}
  \bibinfo{volume}{104} (\bibinfo{year}{2020}), \bibinfo{pages}{106160}.
\newblock
\showISSN{0747-5632}
\urldef\tempurl%
\url{https://doi.org/10.1016/j.chb.2019.106160}
\showDOI{\tempurl}


\bibitem[\protect\citeauthoryear{{de Vries}, Gensler, and Leeflang}{{de Vries}
  et~al\mbox{.}}{2012}]%
        {devries2012}
\bibfield{author}{\bibinfo{person}{Lisette {de Vries}}, \bibinfo{person}{Sonja
  Gensler}, {and} \bibinfo{person}{Peter~S.H. Leeflang}.}
  \bibinfo{year}{2012}\natexlab{}.
\newblock \showarticletitle{Popularity of Brand Posts on Brand Fan Pages: An
  Investigation of the Effects of Social Media Marketing}.
\newblock \bibinfo{journal}{\emph{Journal of Interactive Marketing}}
  \bibinfo{volume}{26}, \bibinfo{number}{2} (\bibinfo{year}{2012}),
  \bibinfo{pages}{83 -- 91}.
\newblock
\showISSN{1094-9968}
\urldef\tempurl%
\url{https://doi.org/10.1016/j.intmar.2012.01.003}
\showDOI{\tempurl}


\bibitem[\protect\citeauthoryear{Gray, Kou, Battles, Hoggatt, and Toombs}{Gray
  et~al\mbox{.}}{2018}]%
        {gray2018}
\bibfield{author}{\bibinfo{person}{Colin~M. Gray}, \bibinfo{person}{Yubo Kou},
  \bibinfo{person}{Bryan Battles}, \bibinfo{person}{Joseph Hoggatt}, {and}
  \bibinfo{person}{Austin~L. Toombs}.} \bibinfo{year}{2018}\natexlab{}.
\newblock \showarticletitle{The Dark (Patterns) Side of UX Design}
  \emph{(\bibinfo{series}{CHI '18})}. \bibinfo{publisher}{Association for
  Computing Machinery}, \bibinfo{address}{New York, NY, USA},
  \bibinfo{pages}{1–14}.
\newblock
\showISBNx{9781450356206}
\urldef\tempurl%
\url{https://doi.org/10.1145/3173574.3174108}
\showDOI{\tempurl}


\bibitem[\protect\citeauthoryear{Grieve, Indian, Witteveen, Tolan, and
  Marrington}{Grieve et~al\mbox{.}}{2013}]%
        {grieve2013face}
\bibfield{author}{\bibinfo{person}{Rachel Grieve}, \bibinfo{person}{Michaelle
  Indian}, \bibinfo{person}{Kate Witteveen}, \bibinfo{person}{G~Anne Tolan},
  {and} \bibinfo{person}{Jessica Marrington}.} \bibinfo{year}{2013}\natexlab{}.
\newblock \showarticletitle{Face-to-face or Facebook: Can social connectedness
  be derived online?}
\newblock \bibinfo{journal}{\emph{Computers in human behavior}}
  \bibinfo{volume}{29}, \bibinfo{number}{3} (\bibinfo{year}{2013}),
  \bibinfo{pages}{604--609}.
\newblock


\bibitem[\protect\citeauthoryear{Kim, Sung, Lee, Choi, and Sung}{Kim
  et~al\mbox{.}}{2016}]%
        {kim2016}
\bibfield{author}{\bibinfo{person}{Dong~Hoo Kim}, \bibinfo{person}{Yoon~Hi
  Sung}, \bibinfo{person}{So~Young Lee}, \bibinfo{person}{Dongwon Choi}, {and}
  \bibinfo{person}{Yongjun Sung}.} \bibinfo{year}{2016}\natexlab{}.
\newblock \showarticletitle{Are you on Timeline or News Feed? The roles of
  Facebook pages and construal level in increasing ad effectiveness}.
\newblock \bibinfo{journal}{\emph{Computers in Human Behavior}}
  \bibinfo{volume}{57} (\bibinfo{year}{2016}), \bibinfo{pages}{312 -- 320}.
\newblock
\showISSN{0747-5632}
\urldef\tempurl%
\url{https://doi.org/10.1016/j.chb.2015.12.031}
\showDOI{\tempurl}


\bibitem[\protect\citeauthoryear{Liu, Gummadi, Krishnamurthy, and Mislove}{Liu
  et~al\mbox{.}}{2011}]%
        {Liu2011}
\bibfield{author}{\bibinfo{person}{Yabing Liu}, \bibinfo{person}{Krishna~P.
  Gummadi}, \bibinfo{person}{Balachander Krishnamurthy}, {and}
  \bibinfo{person}{Alan Mislove}.} \bibinfo{year}{2011}\natexlab{}.
\newblock \showarticletitle{Analyzing Facebook Privacy Settings: User
  Expectations vs. Reality} \emph{(\bibinfo{series}{IMC '11})}.
  \bibinfo{publisher}{Association for Computing Machinery},
  \bibinfo{address}{New York, NY, USA}, \bibinfo{pages}{61–70}.
\newblock
\showISBNx{9781450310130}
\urldef\tempurl%
\url{https://doi.org/10.1145/2068816.2068823}
\showDOI{\tempurl}


\bibitem[\protect\citeauthoryear{Sedghi}{Sedghi}{2014}]%
        {sedghi2014}
\bibfield{author}{\bibinfo{person}{Ami Sedghi}.}
  \bibinfo{year}{2014}\natexlab{}.
\newblock \bibinfo{title}{Facebook: 10 years of social networking, in numbers}.
\newblock
\newblock
\urldef\tempurl%
\url{https://www.theguardian.com/news/datablog/2014/feb/04/facebook-in-numbers-statistics}
\showURL{%
\tempurl}
\newblock
\shownote{(visited on 01/05/2021).}


\bibitem[\protect\citeauthoryear{Shakya, Hughes, Stafford, Christakis, Fowler,
  and Silverman}{Shakya et~al\mbox{.}}{2016}]%
        {shakya2016intimate}
\bibfield{author}{\bibinfo{person}{Holly~B Shakya}, \bibinfo{person}{D~Alex
  Hughes}, \bibinfo{person}{Derek Stafford}, \bibinfo{person}{Nicholas~A
  Christakis}, \bibinfo{person}{James~H Fowler}, {and} \bibinfo{person}{Jay~G
  Silverman}.} \bibinfo{year}{2016}\natexlab{}.
\newblock \showarticletitle{Intimate partner violence norms cluster within
  households: an observational social network study in rural Honduras}.
\newblock \bibinfo{journal}{\emph{BMC public health}} \bibinfo{volume}{16},
  \bibinfo{number}{1} (\bibinfo{year}{2016}), \bibinfo{pages}{233}.
\newblock


\bibitem[\protect\citeauthoryear{Sinclair and Grieve}{Sinclair and
  Grieve}{2017}]%
        {sinclair2017facebook}
\bibfield{author}{\bibinfo{person}{Tara~J Sinclair} {and}
  \bibinfo{person}{Rachel Grieve}.} \bibinfo{year}{2017}\natexlab{}.
\newblock \showarticletitle{Facebook as a source of social connectedness in
  older adults}.
\newblock \bibinfo{journal}{\emph{Computers in Human Behavior}}
  \bibinfo{volume}{66} (\bibinfo{year}{2017}), \bibinfo{pages}{363--369}.
\newblock


\bibitem[\protect\citeauthoryear{Twenge, Joiner, Rogers, and Martin}{Twenge
  et~al\mbox{.}}{2018}]%
        {twenge2018}
\bibfield{author}{\bibinfo{person}{Jean~M. Twenge}, \bibinfo{person}{Thomas~E.
  Joiner}, \bibinfo{person}{Megan~L. Rogers}, {and}
  \bibinfo{person}{Gabrielle~N. Martin}.} \bibinfo{year}{2018}\natexlab{}.
\newblock \showarticletitle{Increases in Depressive Symptoms, Suicide-Related
  Outcomes, and Suicide Rates Among U.S. Adolescents After 2010 and Links to
  Increased New Media Screen Time}.
\newblock \bibinfo{journal}{\emph{Clinical Psychological Science}}
  \bibinfo{volume}{6}, \bibinfo{number}{1} (\bibinfo{year}{2018}),
  \bibinfo{pages}{3--17}.
\newblock
\urldef\tempurl%
\url{https://doi.org/10.1177/2167702617723376}
\showDOI{\tempurl}


\bibitem[\protect\citeauthoryear{Wang, Leon, Scott, Chen, Acquisti, and
  Cranor}{Wang et~al\mbox{.}}{2013}]%
        {wang2013}
\bibfield{author}{\bibinfo{person}{Yang Wang}, \bibinfo{person}{Pedro~Giovanni
  Leon}, \bibinfo{person}{Kevin Scott}, \bibinfo{person}{Xiaoxuan Chen},
  \bibinfo{person}{Alessandro Acquisti}, {and} \bibinfo{person}{Lorrie~Faith
  Cranor}.} \bibinfo{year}{2013}\natexlab{}.
\newblock \showarticletitle{Privacy Nudges for Social Media: An Exploratory
  Facebook Study}. In \bibinfo{booktitle}{\emph{Proceedings of the 22nd
  International Conference on World Wide Web}} (Rio de Janeiro, Brazil)
  \emph{(\bibinfo{series}{WWW '13 Companion})}. \bibinfo{publisher}{Association
  for Computing Machinery}, \bibinfo{address}{New York, NY, USA},
  \bibinfo{pages}{763–770}.
\newblock
\showISBNx{9781450320382}
\urldef\tempurl%
\url{https://doi.org/10.1145/2487788.2488038}
\showDOI{\tempurl}


\bibitem[\protect\citeauthoryear{Wang, Norcie, Komanduri, Acquisti, Leon, and
  Cranor}{Wang et~al\mbox{.}}{2011}]%
        {wang2011}
\bibfield{author}{\bibinfo{person}{Yang Wang}, \bibinfo{person}{Gregory
  Norcie}, \bibinfo{person}{Saranga Komanduri}, \bibinfo{person}{Alessandro
  Acquisti}, \bibinfo{person}{Pedro~Giovanni Leon}, {and}
  \bibinfo{person}{Lorrie~Faith Cranor}.} \bibinfo{year}{2011}\natexlab{}.
\newblock \showarticletitle{"I Regretted the Minute I Pressed Share": A
  Qualitative Study of Regrets on Facebook}. In
  \bibinfo{booktitle}{\emph{Proceedings of the Seventh Symposium on Usable
  Privacy and Security}} (Pittsburgh, Pennsylvania)
  \emph{(\bibinfo{series}{SOUPS '11})}. \bibinfo{publisher}{Association for
  Computing Machinery}, \bibinfo{address}{New York, NY, USA}, Article
  \bibinfo{articleno}{10}, \bibinfo{numpages}{16}~pages.
\newblock
\showISBNx{9781450309110}
\urldef\tempurl%
\url{https://doi.org/10.1145/2078827.2078841}
\showDOI{\tempurl}


\end{thebibliography}

\end{document}